# A COMPARATIVE STUDY ABOVE TWO SELF-TUNING CONTROLLERS WITH APLICATION TO THE CONTROL OF SYNCHRONOUS GENERATOR EXCITATION SYSTEM

Ioan FILIP, Daniel CURIAC, Octavian PROSTEAN, Iosif SZEIDERT

"Politehnica" University of Timisoara, Faculty of Control Engineering and Computer Sciences, Department of Automation and Industrial Informatics
ifilip@aut.utt.ro

Abstract: This paper presents two self-tuning control structures synthesized through the minimization of two criterion functions. It is described the computation methodology of the control laws, both being particularized for the case of the synchronous generator's excitation control. The parameters estimator is considered the recursive least square error (RLSE) algorithm. In order to validate the considered control structures, two comparative study cases by computer simulation are presented.

## 1. INTRODUCTION

Due to the enhanced performances shown by the self-tuning control structures in the case of complex systems placed in a stochastic environment, the following structures presents themselves as a viable and easy implementing alternative in the context of an outstanding development of the computer technology. Although the theoretical basis of the self-tuning controllers are already well known, this control algorithms present themselves as actual solutions (Ljung, 1987) (Wellstead, 1991).

## 2. SYNTHESIS OF THE SELF-TUNING CONTROL STRUCTURES

In the following paragraphs two control structures are presented: self-tuning controller with feedback (error) compensation; self-tuning controller with feedback and reference compensation.

### 2.1 Self-tuning controller with feedback compensation ($J_1$ criterion).

The starting point is represented by the following minimization criterion:

$$(1) \quad J_1 = E\left\{[y(t+1) - w(t)]^2 + [Q'(z^{-1})u(t)]^2\right\}$$

The linearised model of any considered process has the following relation (the last member of the relation is the transfer function of the synchronous generator considered for the following study cases):

$$(2) \quad H(z^{-1}) = z^{-1}\frac{B(z^{-1})}{A(z^{-1})} = z^{-1}\frac{b_3 z^{-3} + b_2 z^{-2} + b_1 z^{-1} + b_0}{a_4 z^{-4} + a_3 z^{-3} + a_2 z^{-2} + a_1 z^{-1} + 1}$$

where: $y(t)$ - process output; $u(t)$ – process input; $e(t)$ – stochastic sequence of independent random variables, of zero average and $s^2$ dispersion (white noise); $d$ – steady state regime process output (for a zero input); $z^{-1}$ – one step delay operator, and

$$C(z^{-1}) = 1 + c_1 z^{-1} + c_2 z^{-2} + c_3 z^{-3} + c_4 z^{-4}$$

a stabile polynomial (noise filter). Minimizing the criterion function described by (1) and considering

$$Q(q^{-1}) = \frac{Q'(0)Q'(z^{-1})}{b_0}$$

we obtain the following result:

$$(3)\quad u(t) = \frac{C(z^{-1})w(t) - F(z^{-1})y(t) - d}{B(z^{-1}) + Q(z^{-1})C(z^{-1})}$$

where: $F(z^{-1}) = z[C(z^{-1}) - A(z^{-1})]$.

Now, if $C(z^{-1}) = 1$ and $d=0$ we obtain $F(z^{-1}) = z[1 - A(z^{-1})]$ and

$$(4)\quad u(t) = \frac{w(t) - z[1 - A(z^{-1})]y(t)}{B(z^{-1}) + Q(z^{-1})}$$

Taking into account that the $A(z^{-1})$ and $B(z^{-1})$ (also for $F(z^{-1})$) polynomial's parameter that occur in control law (relations (3) and (4)) are practically estimations of the real process parameters, the control law can be depicted as follows:

$$(5)\quad u(t) = \frac{w(t) - z[1 - \widehat{A}(z^{-1})]y(t)}{\widehat{B}(z^{-1}) + Q(z^{-1})}$$

where ^ mark the estimations (Fig.1).

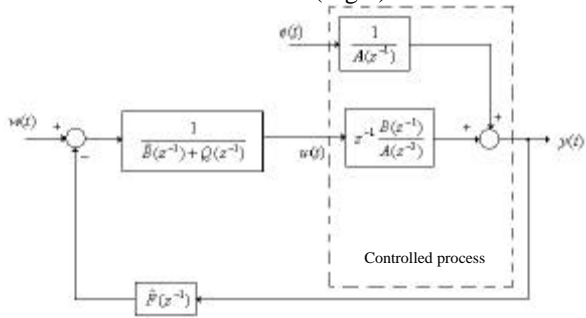

Fig.1. The generalized control structure of the adaptive system based on the $J_1$ criterion ($d=0$, $C(z^{-1})=1$)

Adapting the control laws for the particular expressions of the $A(z^{-1})$ si $B(z^{-1})$ polynomial leads to the following calculus of the adaptive self-tuning command, specific to the considered process:

$$(6)\quad u(t) = \frac{w(t) + \hat{a}_1 y(t) + \hat{a}_2 z^{-1} y(t) + \hat{a}_3 z^{-2} y(t) + \hat{a}_4 z^{-3} y(t)}{\hat{b}_0 + \hat{b}_1 z^{-1} + \hat{b}_2 z^{-2} + \hat{b}_3 z^{-3} + Q(z^{-1})}$$

A general form adopted for the $Q(z^{-1})$ polynomial is $Q(z^{-1}) = r(1 - z^{-1})$ (Xia, 1983)

2.2 *Self-tuning controller with feedback and reference compensation ($J_2$ criterion)*

In this case the criterion function to be minimized is:

$$(7)\quad J_2 = E\left\{[y(t+1) - w(t)]^2 + [Q'(z^{-1})[u(t) - u_r(t)]]^2\right\}$$

where: $w(t)$ - reference input; $u_r(t)$ - steady state regime command.

Similarly with the previous calculus methodology results:

$$(8)\quad u(t) = \frac{w(t) - F(z^{-1})y(t) + Q(z^{-1})u_r(t)}{B(z^{-1}) + Q(z^{-1})}$$

A convenient choice for the $Q(z^{-1})$ polynomial is $Q(z^{-1}) = r$ (considering a reference compensation that assures already the removal of steady state regime error). This case leads us to the following result:

$$(9)\quad u(t) = \frac{w(t) - z[1 - A(z^{-1})]y(t)}{B(z^{-1}) + r} + \frac{r}{B(z^{-1}) + r} u_r(t)$$

In steady state regime $y(t) = w(t)$, so $A(z^{-1})w(t) = z^{-1}B(z^{-1})u_r(t)$ we obtain:

$$(10)\quad u_r(t) = z \frac{A(z^{-1})}{B(z^{-1})} w(t)$$

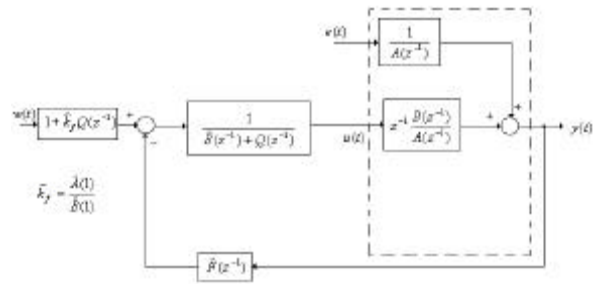

Fig.2. The generalized control structure of the adaptive system based on the $J_2$ criterion ($d=0$, $C(z^{-1})=1$)

By noting $k_f = \frac{A(1)}{B(1)}$, in steady state regime ($z=1$) we obtain: $u_r(t) = k_f w(t)$. If w(t)=ct. results $u_r(t)$=ct., where $1/k_f$ is the process gain coefficient in steady state regime. The proposed solution is $\widehat{k}_f = \frac{\widehat{A}(1)}{\widehat{B}(1)}$ (the permanent estimation of the $\widehat{k}_f$ coefficient on the basis of process parameters estimations). This solution is valid also in the case of a time variable reference (Filip, 1997)

Similarly with the previously case, taking into account that the parameters that occur in the control law are practically estimations of the process's parameters, the control law can be written as follows (for $Q(z^{-1})=r$):

$$(11) \quad u(t) = \frac{-z[1-\hat{A}(z^{-1})]y(t)}{\hat{B}(z^{-1})+r} + \frac{1+r\,\hat{k}_f}{\hat{B}(z^{-1})+r}w(t)$$

where ^ mark the estimations.

We define:

$$(12) \quad \hat{k}_c \stackrel{\Delta}{=} 1 + r\,\hat{k}_f = 1 + r\,\frac{\hat{A}(1)}{\hat{B}(1)}$$

where: $\hat{k}_c$ is the reference compensation parameter.

This parameter, as we can notice from figure 2, assures a reference ($w(t)$) compensation in order to remove any possible steady state regime error. For the considered process we have:

$$(13) \quad u(t) = \frac{\hat{a}_1 + \hat{a}_2 z^{-1} + \hat{a}_3 z^{-2} + \hat{a}_4 z^{-3}}{(\hat{b}_0 + r) + \hat{b}_1 z^{-1} + \hat{b}_2 z^{-2} + \hat{b}_3 z^{-3}} y(t) +$$
$$+ \frac{1+r\hat{k}_f}{(\hat{b}_0 + r) + \hat{b}_1 z^{-1} + \hat{b}_2 z^{-2} + \hat{b}_3 z^{-3}} w(t)$$

and

$$(14) \quad \hat{k}_c = 1 + r\,\hat{k}_f = 1 + r\,\frac{\hat{a}_1 + \hat{a}_2 + \hat{a}_3 + \hat{a}_4}{\hat{b}_0 + \hat{b}_1 + \hat{b}_2 + \hat{b}_3}$$

### 3. STUDY CASES

*3.1. The case of self-tuning control structure based on the minimization of the $J_1$ criterion*

Simulation conditions: the torque has a 0.2 [relative units] step deviation (active load); the RLSE estimator has a forgetting factor of $l = 0.998$; the process is perturbed by a stochastic noise of zero average and $s^2 = 10^{-8}$ variance; penalty factor is $r = 0.01$ (with internal integrative component).

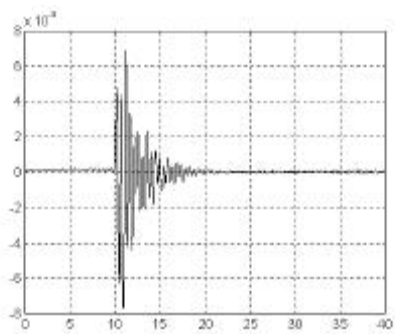

Fig. 3.a. Output voltage (controlled output)

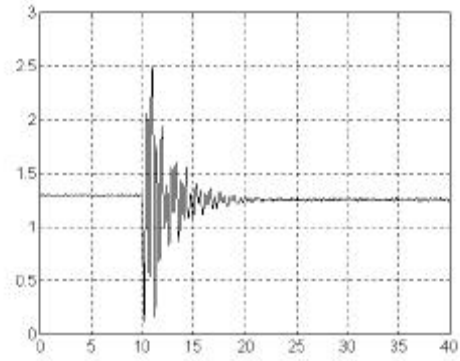

Fig. 3.b. Controller output

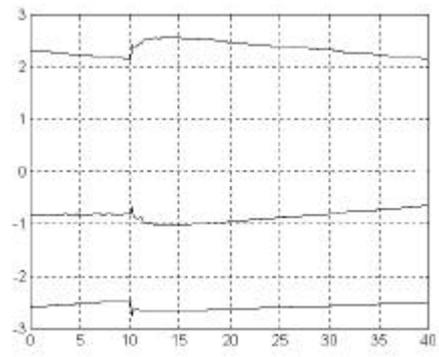

Fig. 3.c. $A(z^{-1})$ polynomial estimated parameters

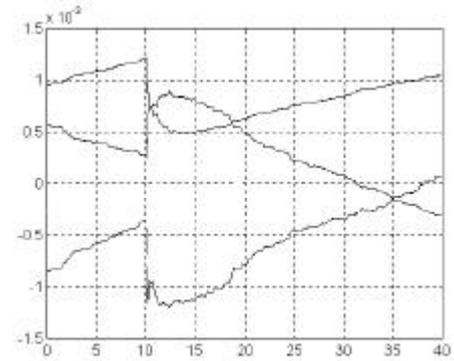

Fig. 3.d. $B(z^{-1})$ polynomial estimated parameters

The command variable (of excitation), presented in figure 3.b. shows a quite large variation. In figure 3.a. can be noticed a good performance of the control structure. All graphical results presented in this paper have time (sec) on the horizontal axis.

*3.2 The case of self-tuning control structure based on the minimization of the $J_2$ criterion*

The simulation conditions are identical with the previous case.

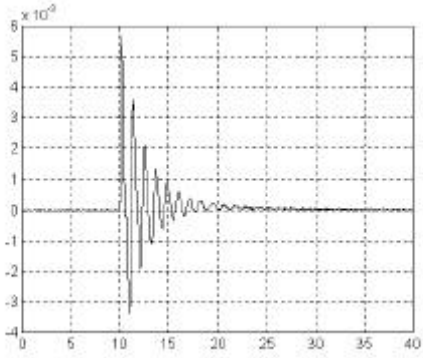

Fig. 4.a. Output voltage (controlled output)

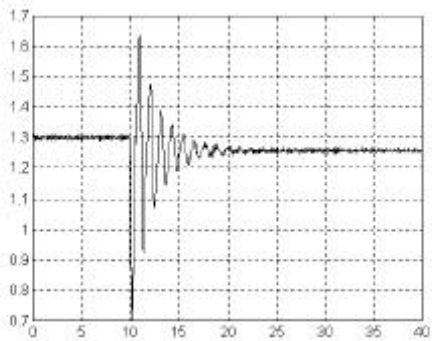

Fig. 4.b. Controller output

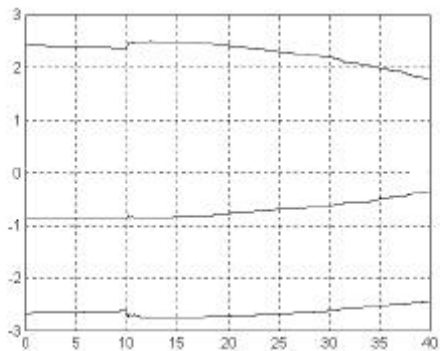

Fig. 4.c. $A(z^{-1})$ polynomial estimated parameters

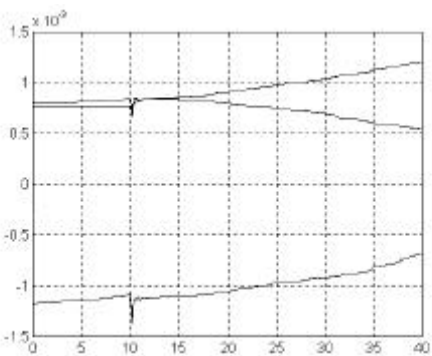

Fig. 4.d. $B(z^{-1})$ polynomial estimated parameters

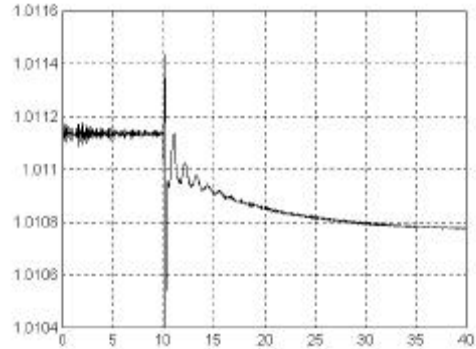

Fig. 4.e. The reference compensation parameter

We can notice that in the case of a step variation of the mechanical torque the performance of the control structure is good (Figures 4.a,b). By considering a forgetting factor of $l = 0.995$ we obtain a evolution of the reference compensation parameter as shown in figure 4.e. This figure highlights a relative slowly evolution to the steady state regime value, due to the reduced estimator's dynamic.

The command variable variance is significantly reduced in compare with the previous case study. (Figure 4.b.). The figures 4.c,d. present the evolution of the estimated parameters, where can be noticed a different evolution in compare to the previous case.

## 4. CONCLUSION

The conducted studies show that both presented self-tuning control structures assure good performances, considering the evolution of the controlled output. (the synchronous generator output voltage). Further, the reference compensation structure assures a smaller command variance (the excitation command voltage). Both control structures present similar performances, even in the condition of different variations of the estimated parameters in closed loop.